\documentclass[doublecol]{epl2}

\usepackage{amssymb}
\usepackage{amssymb}
\usepackage{graphicx}
\usepackage{dcolumn}
\usepackage{bm}
\usepackage{dsfont}
\usepackage{epsfig}

\newcommand{\atanh  }{{\rm{atanh}}}

\newcommand{\bes     }{\mbox{\boldmath$s$}}

\newcommand{\brho    }{\mbox{\boldmath$\rho$}}
\newcommand{\bsigma    }{\mbox{\boldmath$\sigma$}}
\newcommand{\btau    }{\mbox{\boldmath$\tau$}}

\usepackage{color}

\title{Statistical mechanics of LDPC codes on channels with memory}
\author{Izaak Neri\inst{1}\thanks{E-mail: \email{izaak.neri@fys.kuleuven.be}} \and Nikos S. Skantzos\inst{2}
\thanks{E-mail: \email{nikolaos.skantzos@fortis.com}}}
\shortauthor{I. Neri and N. S. Skantzos}

\institute{
\inst{1} Instituut voor Theoretische Fysica, Katholieke Universiteit Leuven, Celestijnenlaan 200D, Leuven B-3001, Belgium\\
\inst{2} BNP Paribas -- Fortis, Waranderberg 3, Brussels B-1000, Belgium
}

\pacs{89.70.-a}{Information and Communication Theory}
\pacs{75.10.Nr}{Spin-glass and other random models}

\abstract{ We present an analytic method of assessing the
typical performance of low-density parity-check codes on
finite-state Markov channels. We show that this problem is similar
to a spin-glass model on a `small-world' lattice. We apply our
methodology to binary-symmetric and binary-asymmetric channels and
we provide the critical noise levels for different degrees of
channel symmetry. }

\begin{document}

\maketitle

\section{Introduction}

A common problem in modern mobile telecommunication systems is that
the strength of the signal varies over time as a result of e.g.\@
the motion of the receiver with respect to the source and the
varying number of obstacles that shadow the signal over time.
Channels describing communication of attenuated signals are termed
`fading channels'.  Fading channels are modeled by finite-state
Markov channels (FSMC)  \cite{Wang1995}. These channels have fueled
significant research activity (for a recent review on the subject
see \cite{Sadeghi}). In FSMCs there exist a number of different
channel states that correspond to the various possible attenuation
factors.  Each of the states describes a memoryless channel
characterized by an error probability, while, the transition from
one state to another occurs according to a stationary Markov
process.  Since there are different states in the FMSC the
error-probabilities between subsequent uses of the channel are
correlated, i.e. there is memory in the channel.

One of the central problems in the domain of error-correcting codes
is the design of codes that reach Shannon's limit. The gap between
the Shannon limit and the computational limit was closed by turbo
codes \cite{Berrou1993} and by low-density parity-check codes (LDPC)
\cite{Gallager63,Mackay1996}. For erasure channels it was shown that
LDPC can reach the Shannon capacity \cite{Luby2001} while for
general symmetric channels one can approach the Shannon limit
\cite{Rich2001}.  To design capacity approaching LDPC-codes one uses
the density evolution (DE) equations to determine the decoding
thresholds \cite{Rich2001T}.  Since channels with memory have a
higher capacity \cite{Mushkin1989, gold} one would like to introduce
memory in the decoding process. Important therefore are the
extensions of turbo codes and LDPC codes to FSMCs \cite{Garcia2002,
Eckford2005}.

Statistical physics has entered the stage of error correcting codes
after the discovery that the decoding problem describing
interactions between parity checks and codeword variables can be
mapped to large frustrated systems of interacting particles
\cite{Sourlas1989}. Since then, physicists have analyzed the
performance of Gallager, MacKay-Neal and Turbo codes over
binary-symmetric, -asymmetric, or real-valued channels
\cite{Vic1999, Mon2000, murayama, Tanaka, Neri2008} (for a review
see \cite{Kabashima2004}). The main actor in this approach is the
generating function of the {\it a posteriori} probability
distribution of codewords. This is similar to the free energy of
spin models. Using the replica method one derives directly the
so-called density evolution equations \cite{Rich2001T}  from the
free energy. Moreover the tools of statistical mechanics can be used
to calculate the error-exponents \cite{skantzos2, rivoire}, MAP-thresholds
\cite{Montanari2005} and modified schemes of belief-propagation
using replica symmetry-breaking effects \cite{Migl2005}. Generally,
the lion's share of the volume of research on error-correcting codes
has been dedicated to memoryless channels. Apart from the work of
\cite{Chert2009}, channels with memory, or any other FSMC models,
have never been to our knowledge analyzed within statistical
physics.

Our work is based on techniques that were developed to analyze
macroscopic properties of `small-world' networks. These systems, due
to their close relation with real-world networks, have been the
subject of intense study from a variety of scientific disciplines
\cite{WattsStro,Dorogovtsev2003,Barabasi2002}. Small-world lattices
have a particular architecture that allows both a high clustering
coefficient and a small shortest path-length (unlike the random
Erd\"os-Renyi graphs). They are constructed by superimposing random
and sparse graphs with a finite average connectivity onto a
one-dimensional ring.  An exact analysis of the thermodynamic
properties of such systems can be found in \cite{Nik2004}. As it is,
FSMCs can be mapped to small-world lattices, whereby messages
between parity checks and codeword-nodes propagate along the sparse
graph while messages between channel-state nodes propagate along the
one-dimensional chain.

In this letter, we present a general method to derive the density
evolution equations for symmetric or asymmetric FSMCs. This includes
an exact analysis of the Gilbert-Elliot channel (GEC)
\cite{Gilbert1960,Elliot1963}. Fully asymmetric cases could be used
to describe burst errors in VLSI circuits \cite{Prad, Blaum}. We
compute the decoding thresholds for the different channels. For
symmetric FSMCs we compare the results to \cite{Eckford2005} while
for memoryless channels to \cite{Wang2005,Neri2008}.

\section{Definitions}
Let us now be more particular. A signal $\bsigma^0\in\{-1,1\}^N$,
prior to its communication over the channel, is encoded to
$\bsigma\in\{-1,1\}^M$ with $M>N$. The set of codewords
$\mathcal{C}$ of an LDPC-code is defined by its parity check matrix
$\mathbb{H}$ through: $\mathcal{C} =
\{\bsigma\in\{-1,1\}^M|\mathbb{H}\ast\bsigma=1\}$ with
$(\mathbb{H}\ast\bsigma)_i =
\prod_{j=1}^M\sigma_j^{\mathbb{H}_{ij}}$ for all  $i=1,\ldots,M-N$.
For $(C,K)$-regular LDPC-codes the parity check matrices are random,
sparse matrices of dimension $(M-N)\times M$ with
$\mathbb{H}_{ij}\in\{0,1\}$ and with $K$ non-zero elements per row
and $C$ non-zero elements per column.

Channel noise can be modeled with the transformation
$\bsigma\to\brho$ where the output of the channel
$\brho\in\{-1,1\}^M$ depends on the input through the state variable
$\bes \in \mathcal{S}^M$:
\begin{eqnarray}
P(\brho|\bes, \bsigma) = \prod^M_{n=1}\Big(P_{\rm chan}(\rho_n|\sigma_n,
s_n)\Big) P_{\rm state}(\bes) \:. \label{eq:probrho}
\end{eqnarray}
The probability of the states $P_{\rm state}(\bes)$ follows a Markov process
\begin{eqnarray}
P_{\rm state}(\bes) =P_{\rm state}(s_1)
\prod^M_{n=1}\mathcal{W}(s_{n+1}|s_n) \:.
\end{eqnarray}

\begin{figure}[t]
\begin{center}
 \epsfig{figure=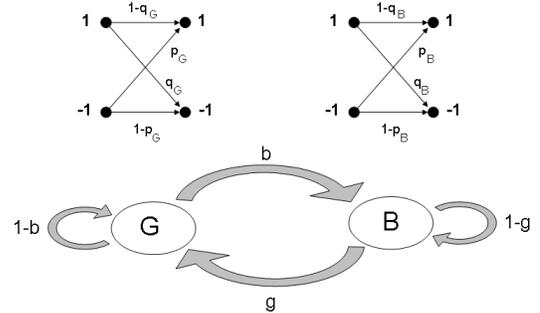,scale=0.35, angle=0}\caption{A graphical representation of the
2-state FSMC where in each state the channel is a binary asymmetric
channel.  The bad state $B$ has a higher noise level than the good
state $G$.} \label{fig:Channel}
\end{center}
\end{figure}

We will denote by $\bsigma^0,\bes^0\in\{-1,1\}^M$ the \emph{true}
codeword and \emph{true} channel state vectors respectively that
were realized during the signal  communication. Depending on the
definition of the Markov process and the channel noise one has
different FSMCs. The derivation of the DE equations stays mainly the
same.   We consider two-state Markov-modulated binary channels. For
these channels the noise is a random variable drawn from the
distribution
\begin{eqnarray}
 P_{\rm chan}(\rho_i=-\sigma_i|s_i, \sigma_i) =
 \left\{ \begin{array}{ccc}q_B,&  & s_i = B,\ \sigma_i = 1  \\  p_B,&  & s_i = B, \ \sigma_i = -1  \\ q_G, && s_i = G, \ \sigma_i = 1\\  p_G, && s_i = G, \ \sigma_i = -1 \end{array} \right.
 \label{eq:chanNoise}
\end{eqnarray}
The channel can be in two states: $\mathcal{S} = \left\{G,
B\right\}$.  Since we take $(p_B+q_B)> (p_G+q_G)$, $B$ is called the
bad state and $G$ is called the good state. The Markov process is
determined by the transition probability $\mathcal{W}$ given by
\begin{eqnarray}
\mathcal{W} \equiv \left[ \begin{array}{cc} 1-b & b \\g &
1-g\end{array}\right] \:, \label{eq:WDef}
\end{eqnarray}
with $g$ the transition probability from state $B$ to $G$ and $b$
the transition probability from state $G$ to $B$ (fig.
\ref{fig:Channel}). We define the memory $\mu_\ell$ at time step
$\ell$ of the Markov process as
\begin{eqnarray}
 \mu_\ell \equiv \mathcal{W}\left[s_\ell = s|s_0 = s\right] - \mathcal{W}\left[s_\ell = s|s_0 \neq s\right]
\end{eqnarray}

From (\ref{eq:WDef}) we find  $\mu_\ell =
\left(1-g-b\right)^\ell\equiv \mu^\ell$ with the time index $\ell =
1, 2, \ldots$ and $\mu\in[-1,1]$. For $\mu>0$ we have persistent
memory: the probability of remaining in a given state is higher than
the steady-state probability of being in that state. For $\mu<0$ we
have an oscillatory memory.   We also define the {\it good-to-bad
ratio} $\rho = \frac{g}{b}$.  The FSMCs we consider are determined
by the 6-tuple $\mathcal{T} = (\mu, \rho, p_B, q_B, p_G, q_G)$.  The
GEC \cite{Gilbert1960,Elliot1963} corresponds to the subset of
channels $\mathcal{T}_{\rm{GEC}} = (\mu, \rho, p_B, p_B, p_G, p_G)$.
We will also consider channels $\mathcal{T}_{\rm{AS}} = (\mu, \rho,
\kappa q_B, q_B, \kappa q_G, q_G)$ with $\kappa\in [0,1]$ and
$\mathcal{T}_{\rm{Z}} = (\mu, \rho, 0, q, q, 0)$.  The latter
channel could be useful for modeling blocks of bad memory or bursts
of unidirectional noise in VLSI circuits.

\section{Density Evolution Equations}
The starting point for the derivation of the DE equations is the
calculation of the generating function $f$ of the {\it a posteriori} probability
distribution of the codeword $\bsigma$ given the channel's output
$\brho$ and the parity check matrix $\mathbb{H}$:
\begin{eqnarray}
 f(\brho, \mathbb{H}) \equiv -\lim_{M\rightarrow \infty} M^{-1}\log \sum_{\bsigma}P(\bsigma|\brho,\mathbb{H}) \:.
\end{eqnarray}
Using Bayes' law and (\ref{eq:probrho}) we obtain
\begin{eqnarray}
\lefteqn{P(\bsigma| \brho, \mathbb{H}) =\frac{P(\brho|\bsigma) P(\bsigma|\mathbb{H})}{P(\brho|\mathbb{H})}}\nonumber
 \\
&&= Z^{-1}p_{\rm init}(\bsigma)
\delta_{\mathbb{H}}\left[\bsigma\right]\sum_{\bes} P_{\rm
state}(\bes) P_{\rm chan}(\brho|\bes, \bsigma)
\end{eqnarray}
with $P(\bsigma|\mathbb{H})$ the initial probability distribution of
the codewords and $Z$ a normalisation constant. We will consider
unbiased sources of i.i.d.r.v: $p_{\rm init}(\bsigma) = 2^{-M}$.
$P_{\rm chan}(\brho|\bes, \bsigma)$ gives the {\it a priori}
probability distribution of the output $\brho$ given the state
vector $\bes$ and the codeword $\bsigma$ (\ref{eq:chanNoise}). The
Kronecker delta constrains the summation only to those codewords
that obey the parity check equation.

Averaging the generating function over the ensemble of parity-check
matrices, true-states, true-codewords and outputs gives
\begin{eqnarray}
 \lefteqn{\hspace{-0mm}-\overline{f} = \lim_{M\to\infty}\frac{1}{M}\sum_{\mathbb{H},\bes^0,\bsigma^0,\brho}P(\mathbb{H})
P_{\rm state}(\bes^0)P_{\rm chan}(\brho|\bes^0,
\bsigma^0)
\nonumber }\\
&& \times\delta_{\mathbb{H}}\left[\bsigma^0\right]\log\left(\sum_{\bsigma, \bes}P_{\rm
state}(\bes)\delta_{\mathbb{H}}\left[\bsigma\right]P_{\rm
chan}(\brho|\bes, \bsigma)\right) \nonumber
\end{eqnarray}
plus irrelevant constant terms. The probability distribution of the parity-check matrices
$P(\mathbb{H})$ of a $(C,K)$-regular code can be written in terms of a
tensor with $K$ indices and elements in $\{0,1\}$, such that the
probability that an element of the tensor is 1 is
$C\frac{(K-1)!}{M^{K-1}}$ and the sum of the elements equals $C$ for
all of its indices (see e.g. \cite{Kabashima2004,Neri2008}). The
free energy $\overline{f}$ can then be calculated using the replica
trick $\langle \log Z\rangle= \lim_{n\to 0}\frac{1}{n}\log
\langle{Z^n}\rangle$.  This results, for $M\rightarrow \infty$, in a
saddle point integral. The free energy at the saddle point is given
by
\begin{equation}
-\overline{f} =\lim_{n\to 0}\frac1n {\rm extr}_{P,\hat{P}}
\Psi\left(P(\bsigma,\sigma),\hat{P}(\bsigma,\sigma)\right) \:,
\label{eq:Extr}
\end{equation}
with $\Psi$ the exponent of the saddle point integral. The
extremization is taken over the order parameter functions
$P(\bsigma,\sigma)$ and $\hat{P}(\bsigma,\sigma)$. These represent
the usual order parameter functions describing finite connectivity
systems, see for instance \cite{Mon1998}, with
$\bsigma=(\sigma^1,\ldots,\sigma^n)\in\{-1,1\}^n$ originating from
the replication of the dynamic codeword-variables while
$\sigma\in\{-1,1\}$ stems from the inclusion of the quenched true
codeword in the order function.

The exponent $\Psi$ reaches a minimum at the values
$\left(P(\bsigma, \sigma), \hat{P}(\bsigma, \sigma)\right)$ that
satisfy the saddle point equations:
\begin{eqnarray}
  \hat{P}(\bsigma, \sigma) &=& \sum_{(\bsigma_1, \sigma_1),\cdots, (\bsigma_{K-1}, \sigma_{K-1})}\prod^{K-1}_{r=1}P(\bsigma_r, \sigma_r)
\nonumber \\
&& \hspace{-20mm}\times\delta\left(\sigma_1\sigma_2 \cdots
\sigma_{K-1}\sigma,
1\right)\prod^n_{\alpha=1}\delta\left(\sigma^{\alpha}_1\sigma^{\alpha}_2
\cdots \sigma^{\alpha}_{K-1}\sigma^\alpha, 1\right)
\label{eq:HPSelfcAsym}
\\
 P(\bsigma, \sigma) &=& \frac{{\rm{Tr}}\left[
V^{N-1}\left(\hat{P}\right) Q\left(\bsigma, \sigma;
\hat{P}\right)\right]}{{\rm{Tr}} \left[
V^N\left(\hat{P}\right)\right]} \label{eq:PSelfcAsym}
\end{eqnarray}
where we defined
\begin{eqnarray}
  \lefteqn{\langle \bes, s^0|Q(\btau, \tau; \hat{P})|\bes', \left(s^0\right)'\rangle =
  }\nonumber
  \\
  &&
  \left(\hat{P}(\btau, \tau)\right)^{C-1} \mathcal{W}\left[\left(s'\right)^0|
  s^0\right] \prod_\alpha \mathcal{W}\left[\left(s'\right)^\alpha|s^\alpha\right] \nonumber
  \\
  && \times\langle \prod_\alpha P_{\rm chan}\left(\rho|s^{\alpha}, \tau^\alpha\right)\rangle_{\rho|s^0, \tau}  \label{eq:QAsym}
\end{eqnarray}
and we introduced the average $\langle\cdot\rangle_{\rho|s^0, \tau}$
over $P_{\rm{chan}}(\rho|s^0, \tau)$. Note that while the summations
over the replicated codeword variables
$\left\{\bsigma_i\right\}_{i=1\ldots N}$ have been performed by
reducing the graph into a single-site problem, the summations over
the replicated channel-state variables
$\left\{\bes_i\right\}_{i=1\dots N}$ is written as a trace over
matrix products in (\ref{eq:PSelfcAsym}). This constitutes the key
difficulty in our problem as we are dealing with the $(2^n+1)\times
(2^n+1)$ replicated transfer matrix:
\begin{eqnarray}
\lefteqn{\hspace{-10mm}\langle \bes, s^0|V(\hat{P})|\bes',
 \left(s^0\right)'\rangle=
  \sum_{\bsigma, \sigma}\mathcal{W}\left[\left(s'\right)^0| s^0\right]
  \prod_\alpha \mathcal{W}\left[\left(s'\right)^\alpha|s^\alpha\right]\nonumber
  }
  \\
   & &
 \times \left(\hat{P}(\bsigma, \sigma)\right)^C\nonumber \langle\prod_\alpha P_{\rm chan}(\rho|s^{\alpha}, \sigma^{\alpha})\rangle_{\rho|s^0, \sigma}
 \end{eqnarray}
To proceed further we now have to make an assumption with regards to
the structure of the replica space. The simplest, replica symmetric
ansatz, assumes that
\begin{eqnarray}
P(\bsigma, \sigma) =2^{-\frac{1}{K}}\int dh W(h|\sigma) \prod_\alpha \frac{e^{h \sigma^\alpha}}{2\cosh\left(h\right)} \:, \label{eq:PAsym}\\
\hat{P}(\bsigma, \sigma) = 2^{-\frac{K-1}{K}}\int du
Z\left(u|\sigma\right)  \prod_\alpha
\frac{e^{u\sigma^\alpha}}{2\cosh\left(u\right)} \label{eq:HPAsym}
\end{eqnarray}
for some densities $W,Z$. For the left- and right-eigenvectors
$L(\bes, s)$, $R(\bes,s)$ of $V$ we now assume
\begin{eqnarray}
 \langle\bes', s'|R\rangle &=& \sum_{s'}\mathcal{P}_R(s')\int dx \Phi_R(x|s')e^{x\sum_\alpha \left(s'\right)^{\alpha}} \\
 \langle L|\bes, s\rangle&=& \sum_{s}\mathcal{P}_L(s)\int dy
\Phi_L(y|s)e^{y\sum_\alpha s^{\alpha}}
\end{eqnarray}
The form of the above two equations follows the central assumption
of \cite{Nik2004,NikCool}. It allows us to take the remaining trace
in (\ref{eq:PSelfcAsym}). All distributions above are normalized at
$n\rightarrow 0$. The densities $\mathcal{P}_R$ and $\mathcal{P}_L$
represent respectively the right- and left- eigenvectors of
$\mathcal{W}$:
\begin{eqnarray}
\mathcal{P}_R(s_0) =
\sum_{s'_0}\mathcal{W}\left[s'_0|s_0\right]\mathcal{P}_R(s'_0)
\\
\mathcal{P}_L(s'_0) =
\sum_{s_0}\mathcal{W}\left[s'_0|s_0\right]\mathcal{P}_L(s_0)
\end{eqnarray}
Following similar computations as in \cite{Nik2004,BolleHeylSk}, we
derive in the limit $n\to 0$ the closed, self-consistent equations
\begin{eqnarray}
  \lefteqn{W(h|\sigma) =  \int \left(\prod^{C-1}_{r=1}du_r
  Z(u_r|\sigma)\right)} \nonumber
\\
&& \times  \int d\zeta M(\zeta|\sigma)\:  \delta\left[h - \zeta -
\sum^{C-1}_{r=1}u_r\right] \label{eq:DensW}
\end{eqnarray}
\begin{eqnarray}
  F(\xi|\sigma) = \int \left(\prod^{C}_{r=1}du_r Z(u_r|\sigma)\right) \delta\left[\xi - \sum^C_{r=1}u_r\right] \label{eq:DensF}
  \end{eqnarray}
and also

\begin{widetext}
\begin{eqnarray}
M(\zeta|\sigma) &=& 2\sum_{s_0,
s'_0}\mathcal{P}_R\left((s')^0\right)\mathcal{W}\left[\left(s'\right)^0|
s^0\right]\mathcal{P}_L(s^0) \int dx dy\Phi_L(y|s_0) \Phi_R(x|s'_0)
\sum_{\rho} P_{\rm chan}(\rho |\sigma, s_0)
\nonumber
\\
&& \times \delta\left[\zeta  - \frac{1}{2} \sum_{\tau}\tau\sigma\log\left(\sum_{s
s'}e^{\left(s'x + sy\right)}\mathcal{W}\left[s'|s\right]P_{\rm
chan}\left(\rho|\tau, s\right)\right)\right] \label{eq:DensM}
\\
Z(u|\sigma)&=& \sum_{\sigma_1, \cdots,
\sigma_K-1}\frac{\delta\left(\sigma_1\cdots\sigma_{K-1}\sigma;
1\right)}{2^{K-2}}
    \int \prod^{K-1}_{\ell=1}dh_\ell
  W(h_\ell|\sigma_\ell)\,\delta\left[u - \atanh\prod^{K-1}_{\ell=1} \tanh(h_\ell)\right] \label{eq:DensZ}
\\
  \Phi_R(x|s_0) &=&\sum_{s'_0}\mathcal{W}\left[s'_0|s_0\right]
  \sum_{\tau}\frac{1}{2}\int d\xi F(\xi|\tau) \int dx'
\Phi_R(x'|s'_0)\sum_\rho P_{\rm chan}(\rho| \tau, s_0) \nonumber
\\
&&
  \times\delta\left[x-
\frac{1}{2}\sum_{s}s\log\left(\sum_{s'}\mathcal{W}\left[s'|s\right]\frac{e^{x's'}}{2\cosh\left(x'\right)}\right)\right.
-\left. \frac{1}{2}\sum_{s}s\log\left(
\sum_{\sigma}\prod_r\frac{e^{\xi\sigma\tau}}{2\cosh \xi}P_{\rm
chan}\left(\rho|\sigma, s\right)\right)\right] \label{eq:DensPhiR}
\\
  \Phi_L(x|s'_0) &=&
\sum_{s_0}\frac{\mathcal{W}\left[s'_0|s_0\right]\mathcal{P}_L(s_0)}{\mathcal{P}_L(s'_0)}
  \sum_{\tau}\frac{1}{2}\int d\xi F(\xi|\tau) \int dx'
\Phi_L(x'|s_0)\sum_\rho P_{\rm chan}(\rho|\tau, s_0)
\nonumber \\
&&  \times\delta\left[x-
\frac{1}{2}\sum_{s'}s'\log\left(\sum_{s}\mathcal{W}\left[s'|s\right]\frac{e^{x's}}{2\cosh\left(x'\right)}\left(\sum_{\sigma}\frac{e^{\xi
\sigma\tau}}{2\cosh \xi}P_{\rm chan}\left(\rho|\sigma,
s\right)\right)\right)\right] \label{eq:DensPhiL}
\end{eqnarray}
\end{widetext}

\begin{figure}[h]
\begin{center}
 \epsfig{figure=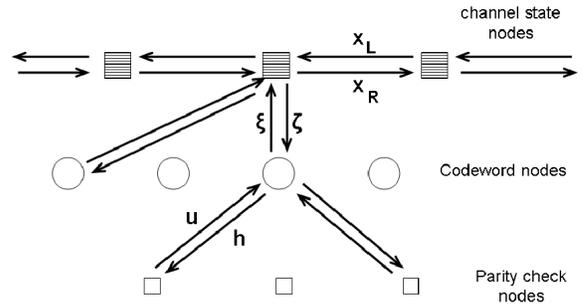,scale=0.35,
angle=0}\caption{The tripartite graph and the messages propagating along the graph for a LDPC code on channels with memory.}\label{fig:messages}
\end{center}
\end{figure}

Equations  (\ref{eq:DensW}-\ref{eq:DensPhiL}) are
the DE equations for the binary asymmetric two-state
Markov channel. They describe the evolution of the densities of
messages propagating along a tripartite graph. The graph consists of
a chain of channel-state nodes connected to codeword nodes and these
in turn to parity check ones. This graphical representation of the
decoding process corresponds to an efficient algorithm
\cite{EckfordThesis}, equivalent to the sum-product algorithm used
in channels without memory.  The tripartite graph has three
different sets of vertices: the set $V_{\rm code}$ of codeword
nodes, the set $V_{\rm pc}$ of parity check nodes and the set
$V_{\rm chan}$ of channel-state nodes, see fig.
\ref{fig:messages}. Due to the presence of memory there are 6 types
of messages propagating according to:

\begin{tabular}{lll}
  Message & From & To \\ \hline
  $h_{i\rightarrow a}$ & $i\in V_{\rm code}$ & $a\in V_{\rm pc}$ \\
 $u_{a\rightarrow i}$ &  $a\in V_{\rm pc} $& $i\in V_{\rm code}$\\
$\zeta_{c\rightarrow i}$ & $c\in V_{\rm chan}$ &$i\in V_{\rm code}$\\
$\xi_{i\rightarrow c}$ & $i\in V_{\rm code}$ & $c\in V_{\rm chan}$\\
$x_{R; c\rightarrow c+1}$ & $c\in V_{\rm chan}$ & \\
$x_{L; c\rightarrow c-1}$& $c\in V_{\rm chan}$&\\
\end{tabular}

The update equations for single-graph instances for these messages
(the so-called `message-passing' equations) \cite{EckfordThesis}
correspond to the functions within the delta functions in the DE
equations (\ref{eq:DensW}-\ref{eq:DensPhiL}).

\begin{figure}[t]
\begin{center}
\epsfig{figure=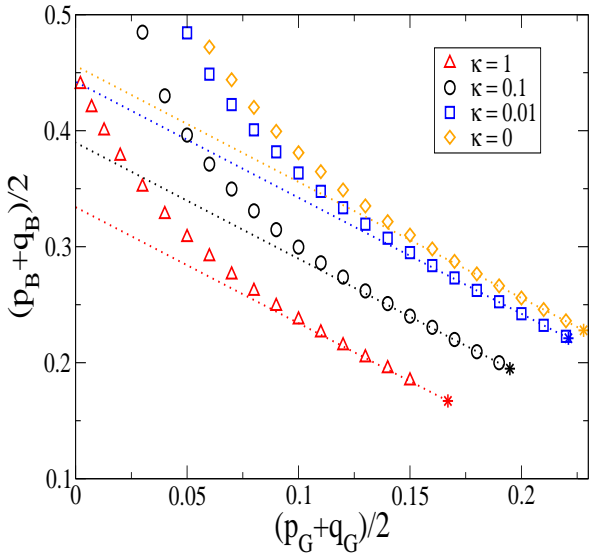,height=\linewidth,
width=\linewidth, angle=270}
\caption{Decoding thresholds (markers) of a $(3,4)$-regular LDPC
code on a  $\mathcal{T}_{\rm AS}$-channel presented in the
space of $\left(\frac12(p_B+q_B),\frac12(p_G+q_G)\right)$ for different
values of the asymmetry $\kappa\in[0,1]$. For all symbols the memory equals
$\mu = 0.90$ while the good-to-bad ratio is $\rho = 1$.  The dotted
line represents the memoryless threshold $\mu=0$. }\label{fig:asym}
\end{center}
\end{figure}

\begin{figure}[t]
\begin{center}
\epsfig{figure=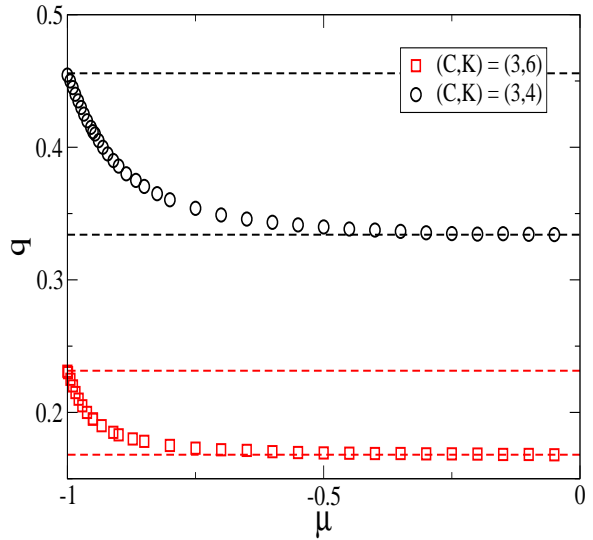,height=\linewidth,
width=\linewidth, angle=270}
\caption{Decoding thresholds in the  $(\mu,q)$ space of
$(C,K)$-regular LDPC codes on a $\mathcal{T}_{\rm Z}$-channel for a
good-to-bad ratio $\rho = 1$.  The upper dashed lines represent the
decoding thresholds for the corresponding memoryless binary
asymmetric channel. The lower one corresponds to a value that is twice that
of the decoding threshold of a memoryless binary symmetric
channel.}\label{fig:Z}
\end{center}

\end{figure}

\section{Results}

We are interested in deriving the critical noise levels beyond which
decoding is not possible. This information can be obtained through
the observable $\rho_\sigma\equiv \frac{1}{|I_\sigma|}\sum_{i\in
I_\sigma} \sigma_i=\int dh \tilde{W}(h|\sigma) {\rm sign}(h)$ where
$I_\sigma$ describes the sublattice $I_\sigma = \{i\in V_{\rm code} |
\sigma_i^0=\sigma\}$ and $\tilde{W}$  the distribution of the
marginals of the decoding variables
\begin{eqnarray}
\lefteqn{\tilde{W}(h|\sigma) =  \int \left(\prod^{C}_{r=1}du_r
Z(u_r|\sigma)\right)} \nonumber
\\
&& \times  \int d\zeta M(\zeta|\sigma)\:  \delta\left[h - \zeta -
\sum^{C}_{r=1}u_r\right] \label{eq:DensTW}
\end{eqnarray}
The value $\rho_\sigma = 1$ corresponds to perfect decoding
(ferromagnetic phase) while $\rho_\sigma<1$ describes decoding
failure (paramagnetic phase). We detect the transition by
numerically solving the DE equations (e.g.\@ through population
dynamics \cite{Mez2001}).

The decoding thresholds in the parameter space
$\big(\frac12(p_G+q_G),\frac12(p_B+q_B)\big)$  for a Gallager
$(C,K)=(3,4)$ code on a $\mathcal{T}_{\rm AS}$-channel are shown in
fig. \ref{fig:asym}. Dotted lines separate ferro- from
paramagnetic solutions for memoryless channels with $\mu=0$, while symbols
correspond to channels with memory for $\mu=0.9$.  We show four
degrees of channel asymmetry characterized by the variable $\kappa =
p_B/q_B = p_G/q_G$. Note that to simplify the presentation of our
results the two channel states have here the same $\kappa$. The decoding thresholds for $\kappa=0$ are computed from the DE equations for $\kappa\rightarrow 0$, which can be derived when rescaling the fields $h\rightarrow \beta h$, $u\rightarrow \beta u$, $\zeta\rightarrow \beta \zeta$ and $\xi\rightarrow \beta \xi$ with $\beta = -\frac{1}{4}\log(\kappa)$.  The
points marked by the star-symbols correspond to the points where the
two channel states have the same error probability, $p_B=p_G$ and are taken from \cite{Neri2008}.
Beyond the star-symbol (lower-right part of the fig.) the roles of
the `good' versus the `bad' channel are interchanged. In this fig.
we also see that both the presence of memory and that of asymmetry
in the channel allows for higher noise levels. In the limiting cases
of  $\kappa=1$ our results agree very well with
those of \cite{Eckford2005}. In
table \ref{table:res} we give the decoding thresholds corresponding
to fig. \ref{fig:asym}.

In fig. \ref{fig:Z} we present results from the channel
$\mathcal{T}_Z$ in which there exist two Z-type states:
$(p_B,q_B)=(0,q)$ and $(p_G,q_G)=(q,0)$ (hence the terms `good' vs
`bad' are not very meaningful here). This type of configuration can
model `burst-error' channels where a very large number of
consecutive bits appear corrupted while the corruption is selective
with regards to the input symbol. We show results in the $(\mu,q)$
space for Gallager $(C,K)=(3,4)$ and $(4,6)$ codes. The lower dashed
line corresponds to the noise level $2q_{{\rm BSC}}$ where $q_{{\rm
BSC}}$ is the critical level of the memoryless binary-symmetric
channel. The fact that the marker at $\mu=0$ coincides with the
dashed line is not a coincidence since in this limit the channel has
two complementary Z-type states without memory, and therefore, with
an equal transition probability between them. The upper dashed line
corresponds to the critical noise level of a memoryless Z-channel
\cite{Neri2008}. At $\mu=-1$ the transition probabilities become
$b=g=1$ and thus the channel oscillates between the two states. We
note that this fig. is symmetric with respect to the $\mu=0$ axis;
a property that also follows from the DE equations.

\begin{table}
\begin{tabular}{ccccc}
 &     $\kappa = 1$ &  $\kappa = 0.1$ & $\kappa = 0.01$ & $\kappa = 0$
\\
 \cline{2-5} $(p_G+q_G)/2$ & \multicolumn{4}{c}{$(p_B+q_B)/2$} \\
 \hline
0.20 & /& 0.190(2)& 0.242(2) & 0.256(2)\\
0.08& 0.262(2)& 0.331(2)&0.401(2)  & 0.420(2)\\
0.05 & 0.308(2)&0.396(2) & 0.484(2) & /\\
\hline
\end{tabular}
\caption{Critical noise levels for the (3,4)-regular LDPC code on a
$\mathcal{T}_{\rm AS}$-channel with memory $\mu = 0.9$, good-to-bad
ratio $\rho = 1$ and for four degrees of channel asymmetry.}
\label{table:res}
\end{table}

\section{Conclusions}

Error-correcting codes on channels with memory are known to
outperform the traditional ones on memoryless channels. They can be
used in modern mobile communication systems or to model burst-error
channels. In this letter we have presented a technique for deriving
the density evolution equations for multi-state channels. This
method is based on the diagonalisation of replicated transfer
matrices that was originally developed to study `small-world'
systems. It turns out that the representation of the LDPC
multi-state decoding problem on graphs shares a common architecture
with `small-world' systems: In memoryless channels, decoding occurs
with message-passing between symbol variables (the `spins') which
are connected to parity-check variables (the `couplings'). Channels
with memory introduce a new element to this hypergraph which can be
seen and treated as a chain of channel-state variables with
nearest-neighbor interactions.

We have presented results for the Gilbert-Elliott channel and its
generalisation to asymmetric two-state channels with memory. The
density evolution equations that follow from the analysis reproduce
very well the special limiting cases of the GEC or the memoryless
binary-asymmetric channel. The method can be applied to a variety of
multi-state error-correcting codes, such as multi-symbol, gaussian-,
non-Markovian or intersymbol-interference channels. From a
statistical physics point of view an interesting future direction
would be the inclusion of replica symmetry-breaking effects
\cite{HatchettWemmenhoveNikolet} which might correct the critical
noise levels we present here.

\acknowledgments
We would like to thank Bastian Wemmenhove who participated in the initial stages of this work.
NS thanks S.d. Guzai for inspiring communication.  IN is grateful to D\'esir\'e Boll\'e for guidance.

\end{document}